# The Light Quanta Modulated Physiological Response of *Brassica Juncea* Seedlings Subjected to Ni(II) Stress#

N. Dasgupta-.Schubert[1]*, S. Alexander[3)], L. Sommer[3)], T. Whelan[4)]

R. Alfaro Cuevas Villanueva[1)], M. E. Mendez Lopez[2)], M. W. Persans[5)]

*1) Instituto de investigaciones Químicas y Biológicas, 2) Facultad de Biología, Universidad Michoacana de San Nicolás de Hidalgo, Ciudad Universitaria, Morelia, Mich., C.P. 58060, Mexico.*

*3) Dept. of Physics, Southwestern University, 1001 E. University, Georgetown, TX, U.S.A.*

4) *Dept. of Chemistry, 5) Dept. of Biology, University of Texas Pan American, 1201 W. University Drive, Edinburg, TX 78539, U.S.A.*

*Author to whom correspondence should be addressed. Email: nita@ifm.umich.mx, phone: +(52) 443 326 5788 Ext. 128; Fax: +(52) 443 326 5790

**Abstract**

This work is a study of the inter-relationship between parameters that principally affect metal up-take in the plant. The relationships between the concentration of metal in the growth medium, $C_s$, the concentration of metal absorbed by the plant, $C_p$, and the total biomass achieved, M, all of which are factors relevant to the efficiency of phytoremediation of the plant, have been investigated via the macro-physiological response of *Brassica juncea* seedlings to Ni(II) stress. The factorial growth experiments treated the Ni(II) concentration in the agar gel and the diurnal light quanta (DLQ) as independently variable parameters. Observations included the evidence of light enhancement of Ni toxicity at the root as well as at the whole plant level, the shoot mass index as a possible indicator of shoot metal sequestration in *B. juncea*, the logarithmic variation of $C_p$ with $C_s$ and the power-law dependence of M on $C_p$. The sum total of these observations indicate that for the metal accumulator *B. juncea* with regard to its capacity to accumulate Ni, the overall metabolic nature of the plant is important – neither rapid biomass increase nor a high metal concentration capability favor the removal of high metal mass from the medium, but rather the plant with the moderate photosynthetically driven biomass growth and moderate metal concentrations demonstrated the ability to remove the maximum mass of metal from the medium. The implications of these observations in the context of the perceived need in phytoremediation engineering to maximize $C_p$ and M simultaneously in the same plant, are discussed.

**Key words**: Bioremediation, heavy metals, phytoremediation.

## 1. Introduction

The ability of certain plants to absorb large quantities of heavy metals from contaminated soil and water, holds significant promise for the deployment of “green technologies” such as phytoremediation towards environmental clean-up [1]. While the technology presents several distinct advantages over conventional remediation methods [2], much still needs to be learnt about the physico-chemical and biological factors that guide the movement and speciation of heavy metals within the rhizosphere and their subsequent up-take, translocation and detoxification by the plant [3]. Of the several technologies that constitute phytoremediation, phytoextraction appears to be optimal for the clean-up of those types of soil where it can be applied, because it is only in this that the toxic metal, absorbed by the root and sent to aerial shoots via xylem transport, can be entirely removed by the subsequent harvest of the shoots [1-5]. While diverse and complex processes specific to the geochemistry of the metallic compound present in the soil and the biochemistry and physiology of metal up-take by the plant, govern the process of phytoextraction at the molecular level [6, 3], these factors ultimately affect the efficiency of phytoextraction at the whole plant level [7, 8], via the mass $M_{metal}$ of heavy metal extracted from the soil per crop. $M_{metal}$ is the product of M the mass of dry plant tissue produced per crop and $C_p$ the concentration of metal in the plant tissue.

$$M_{metal} = M . C_p \qquad (1)$$

The objective of phytoextraction engineering is to maximize $M_{metal}$ for the crop which in the ideal case is served by the simultaneous high values of both M and $C_p$. Phytoremediating plant species generally fall into the categories of high M but relatively low $C_p$ plants, such as maize (*Zea mays*), or the so-called metal hyper-accumulators, e.g. *Thlaspi caerulescens (Brassicaceae)*, that are typically characterized by high $C_p$ but low

M [5]. One aim of plant molecular biologists is to identify novel genes important for phytoremediation including regulatory networks and tissue-specific transporters, and subsequently to manipulate the expression of these genes in high biomass species [3,5,9]. However, the question of the exact trade-off between metal toxicity and plant biomass development as dictated by the whole plant physiological response in quantitative terms, requires more detailed studies. It is a fundamental question that dictates the outcome of any phytoremediation scheme whether using naturally selected species or genetically modified ones. The accuracy of predictive models of phytoextraction efficiency depends on such studies and thereby potentially, the acceptance of this emerging technology by regulatory bodies and enterprises.

The calculation of $M_{metal}$ (eqn. 1) forms the basis of estimates designed to find out how many successive crops would be required to bring about the desired reduction in the soil concentration of the metal [10,7]. Two key assumptions are generally made: (a) that $C_p$, or its analogue the bio-concentration factor, BCF[1], remains constant as the soil metal concentration, $C_s$, changes (BCF is the ratio of $C_p$ to $C_s$) and (b) that M is unaffected by $C_p$ [10,7]

The objective of the present work is to examine the validity of both these assumptions and in doing so to arrive at an empirically quantitative understanding of the relationship between the metal concentration in the medium, $C_s$, its concentration in the plant $C_p$ and the biomass M at the whole plant level. The physiological stress impact of increased metal concentration ($C_p$) within the plant is likely to affect M as well as the fact that the soil-plant system represents two mutually interacting phases in which the metal composition in one phase is likely to affect the other by means of the system's thermodynamics. Moreover the exact nature of the relationships between $C_p$ and $C_s$ and between $C_p$ and M are likely to depend on the metabolic status of the plant and its species.

[1] BCF is generally defined as the ratio of the concentration of metal absorbed by the plant, $C_p$, to the concentration of metal remaining in the soil. For non-hyperaccumulators where the absorbed metal concentration is generally low, it can be shown that this definition is very nearly equivalent to the ratio of $C_p$ to the initial soil concentration of the metal, $C_s$.

We base our study on the detailed examination of several macro-physiological stress indices such as the temporal root length, plant total mass (M) and shoot mass index vis à vis the measured absorbed concentration of the metal ($C_p$ ) in the 2 week old seedlings of the metal accumulator *Brassica Juncea*. This plant is known to phytoextract Ni(II) [11-13] and is a quick-growing relatively large biomass plant that has often been cited as a potential phytoremediant [14,15]. The factorial growth experiments under controlled conditions, involved variable Ni(II) (as nickel acetate tetrahydrate) concentrations ($C_s$) in the agar gel growth medium and variable diurnal light quanta. The latter is used to provide the means of independently varying the biomass via the photosynthetic quantum. It is not used in this work for fundamental studies in photosynthesis. The seedling phase of growth is chosen because metal stress effects on biomass are likely to be the most prominent at this stage, on account of the fast growth rate and the large surface to volume ratio of the root which is the point of entry for the metal ion. The data are analysed in terms of the mathematical functions that best fit the experimental results for $C_s$, $C_p$ and M.

## 2. Materials and Methods

*B. Juncea* seeds (no. 182931, USDA North Central Regional Station, Ames, Iowa 50011-1170, U.S.A) were surface sterilized in a sterile laminar flow-hood. Each batch had ~180 seeds. 60 were germinated and grown in the dark (DARK), 60 at an illumination of 9h of light per day (9H) and 60 at 24h of light per day (24H). The seeds were allowed to germinate in a sterile environment for 48 h then transferred, in the sterile laminar flow-hood, to large (25 cm X 25 cm) sterile petri-dishes containing sterile 1/2xMS (Caisson Laboratories , Rexburg, ID) and MES agar gel growth medium (Sigma-Aldrich, St. Louis, MO) treated with 0 (Control), 50, 100, 150 and 200 μM concentrations of nickel acetate tetrahydrate (Sigma-Aldrich, St. Louis, MO). Each plate contained 12 germinated seeds with each seed being placed at the same pre-marked distance from the top of the plate. The seedlings were grown for a period of ~15 days in the growth chamber under a temperature of 23 $^{0}$C and a relative humidity of 60%. All 5 plates corresponding to the 5 Ni(II) concentrations for each diurnal light period were placed on the same shelf. The DARK plates were completely covered with Al foil. The light intensities on the shelves for the 9H and 24H seedlings were measured with a Bausch&Lomb L1-1000 light meter to be about the same and translated to the diurnal light quanta (DLQ) of 4.32 X $10^{6}$ and 1.20 X $10^{7}$ μmoles photons.$m^{-2}$ $d^{-1}$ respectively. Incremental root lengths for each seedling were measured at 8h intervals. The average cumulative root lengths and their errors (determined propagatively) were calculated for each plate from these incremental lengths.

After 2 weeks growing time, all seedlings from each plate were combined to make one composite sample per plate, dried at 60 $^{0}$C for 48h, weighed and digested according to the procedure given in USEPA SW846 Method 3050. Errors on the masses were computed as the standard errors of the average mass of the seedlings in each plate. The average dry mass of the seedlings in each plate divided by the number of seedlings harvested, gave the dry mass per seedling, M. Samples were then analysed for Ni concentration in the seedlings, $C_p$, using flame atomic absorption spectroscopy, FAAS, (Perkin Elmer AAnalyst 800) using deuterium background correction and a high efficiency nebulizer. Details of the analytical method are provided in [16].

To cross-check the metal concentration analysis on the very low total metal contents of the seedlings, at least one of the sample sets, the 24H samples, was also analysed for Ni using epithermal neutron activation analysis (ENAA) at the research reactor of the University of Texas at Austin (Nuclear Engineering Teaching Laboratory of the J.J. Pickle Research Campus). The liquid samples of digested *B. Juncea* were leak sealed in plastic vials, mounted on the reactor's rotary specimen rack, irradiated for 8h at 950 kW, cooled for 2 months and the product $^{58}Co$ was gamma spectrometrically analysed on a high efficiency low back-ground Compton suppressed high purity Germanium (HPGe) detection system. The data was processed to yield the total Ni concentrations. It was found that the ENAA and the FAAS concentration results mutually agreed within 15%, which serves to substantiate the reliability of the elemental analysis in the present experiments.

Precision values on $C_p$ were computed as the standard errors over experimental replicates. The $M_{Ni}$ were computed using eqn. 1 and the error values were obtained after appropriate propagation over the errors for both M and $C_p$. Nickel concentrations in the control seedlings (where no Ni was added) was below the detection limit of 35 $\mu g\ g^{-1}$ for all 3 three light quanta.

The whole experiment was repeated in triplicate for the plant growth and in duplicate for the FAAS analysis.

## 3. Results and Discussion

Figs 1 and 2 show the rate of root length growth as a function of Ni(II) concentration in the medium, $C_s$, and diurnal light quanta (DLQ). At $C_s \leq 50$ $\mu M$, DLQ enhances root growth after the appearance of the first true leaves by the photosynthetically generated energy as seen for 9H and 24H where the asymptotic growth fade-off in the absence of DLQ, as for DARK, is arrested and a constant linearly increasing growth contribution proportional to the DLQ, sets in. The $C_s = 50$ μM medium was observed to enhance growth over the Control at all stages of the seedling's life. Fig.2 shows the growth fall-off with higher $C_s$ but the trend is almost a "mirror-image" of fig.1: DARK does the best and 24H the worst. At concentrations > 50 μM, light appears to enhance Ni toxicity despite the increased opportunity for photosynthesis.

Fig.3 shows the variation of biomass M per seedling with $C_s$ and DLQ. Masses drop with $C_s$ for all DLQ but at any given $C_s$, are the maximum for the 24H seedling. Again the relative growth enhancement at 50 μM is evident for DARK and 9H but for 24H, M drops monotonically. Also this rate of fall-off with $C_s$ is relatively the highest, indicating a probable light enhancement of Ni toxicity for biomass development, as well.

In the context of phytoextraction, shoot and root mass (or length) would be important with regard to the capacities to compartmentalize metal in aerial shoots and the uptake of the metal. The ratio of shoot mass to root mass/length, referred as the shoot mass index or SMI in this work, would indicate their relative magnitude [17,15]. The root of the *B. juncea* seedling was observed to be principally a single conically elongated primary root with little branching. To know the effect of $C_s$ and DLQ on shoot growth we model the root as an elongated cone. The radius at the root base ($r_{root}$), and root tissue density ($\rho_{root}$), were considered to be the same for all the seedlings and independent of DLQ and $C_s$. Since all the seeds were germinated identically in the absence of metal stress, a constant root base radius seems reasonable. The root mass $M_{root}$ can then be written as,

$$M_{root} = 1/3\,[\pi .\,(r_{root})^2 .\, l_{root} .\, \rho_{root}] = K_{root} .\, l_{root} \quad (2)$$

Where $l_{root}$ and $K_{root}$ are the root length at the end of the two-week growth period and the constant multiplier respectively. Equation 2 implies that the root mass mainly varies with root length. Writing the total mass of the seedling as M and the shoot mass as $M_{shoot}$ ,

$$M = M_{shoot} + M_{root} \quad (3)$$

Or,

$$M_{shoot} / (K_{root} .\, l_{root}) = \{M / (K_{root} .\, l_{root})\} - 1 \quad (4)$$

This means that the ratio of shoot mass to root mass (SMI) is proportional to the ratio of the total mass to the root length since all other factors are constant. The model allows one to estimate SMI on the seedling non-destructively by measuring the root length and total mass only. The model does not hold for densely branched roots.

The variation of SMI with $C_s$ and DLQ are shown in fig 4. SMI generally increases with $C_s$ for both 9H and 24H indicating possibly the tendency of the plant to increase sequestration volume for the Ni(II) by increasing shoot mass. Zavoda et al [15] have shown that Ni is preferentially sequestered in the leaf of *B. juncea* which lends support to this conjecture.

The soluble metal ion enters the root endodermal cells or apoplast by passive and/or active processes [5.9]. The number of metal ions that enter would scale with the root surface area, which in the single conically elongated root would increase with the increase of root length. The relationship between the Ni concentration in the medium and in the plant i.e. between $C_s$ and $C_p$, is shown in fig. 5. DARK has the highest values of $C_p$ for any given $C_s$ which is the combined result of the least suppression of DARK's root length by high Ni concentration in the gel (fig. 3) and its low total mass (fig. 4). For the same reasons, 24H shows the least values of $C_p$. The data were found to follow a logarithmic relationship (eqn 5) (SIGMAPLOT [18]) indicated by the broken lines. The constant $C_0$ and amplitude ‘a’ values for DARK, 9H and 24H were {-3603.78, 981.52}, {-1438.8, 416.93} and {-162.3, 74.44} respectively. The corresponding squares of the correlation coefficient, $R^2$, values at 95% confidence level, were 0.86, 0.96 and 0.95.

$$C_p = C_0 + a \ln(C_s) \qquad (5)$$

This logarithmic approach to saturation is contrary to the assumption of the constancy of plant metal concentrations with changing soil (or growth-medium) metal concentrations or the constancy of the BCF which would imply a constant linear proportionality between $C_p$ and $C_s$ (see Introduction). Infact the non-linear diminution of $C_p$ as $C_s$ increases is expected from the Le Chatelier-Braun principle, applicable to equilibrium as well as non-equilibrium thermodynamics, the latter being more appropriate for living systems [19]. This principle states that a chemical system responds so as to minimize the applied external forces. As the concentration within the plant builds up (increasing $C_p$) the response is to minimize further ingress of metal ions even though the external metal concentration ($C_s$) is not necessarily low, resulting in a tendency towards saturation of the metal ion within the plant, as seen in fig. 5. The implication of eqn 5 is that there will be a limit to the metal extraction capability of the non-senescing plant, in this case *B. juncea*, in a given metal contaminated medium that is dictated by the totality of its bio-physicochemical response to the metal in its environment. It is also interesting to note that the form of eqn. 5 is similar to the equation relating the concentration build-up with time of a product in a first-order chemical reaction which is well-known to result from the action of the Le Chatelier-Braun principle on the chemical equilibrium.

Recently, Rooney et al [6], have presented a detailed and thorough study of the variability of soil geochemistry (e.g. factors such as pH, soil cation exchange capacity CEC etc.) on the availability of soluble Ni(II) to barley and tomato, which was found to be the key factor influencing Ni toxicity to the plants. While the agar growth substrate does not represent true soil, it presents a uniform, homogeneous medium of solubulized Ni(II) distribution. Thereby an unpredictable variability in the experimental results of $C_p$ brought in by unknown Ni(II) retentions in the substrate, are considerably reduced. This is essential if conclusive information linking the two parameters, $C_s$ and $C_p$ is desired. In order to examine a different growth medium but which has the same characteristics of solubulized Ni distribution, we examined the data of the aquaculture experiment of Salt et al [12], wherein the same metal-plant combination (Ni and *B. juncea*) had been used. In this experiment, 5 day old seedlings of *B. juncea* were exposed for 48h to graded concentrations of $Ni(NO_3)_2$ in an aquaculture medium that was constituted with metallic cations and suitable anions, to resemble groundwater. Their results of $C_p$ for the three different $C_s$ values used, are shown in fig. 5. Their results were found to conform well to eqn. 5 ($C_0$ = 96.85, a = 78.41, $R^2$ = 0.97). We notice that the plant Ni concentration in their experiment was lower than the values for DARK and 9H at equivalent growth medium Ni concentrations possibly because of the relatively lower exposure time, the competition for up-take by other cations present in the medium and the early growth phase of the plant. However, these differences are merely in magnitude - the underlying macro-physiological response (eqn. 5) linking $C_p$ to $C_s$ is the same in both studies, as would be expected on the basis of the Le Chatelier thermodynamic principle. In this context it is worth noting that Dasgupta-Schubert et al [20] have recently shown that the logarithmic saturation (eqn. 5) behavior holds for metal up-take by the plant (*B. juncea*) from complex soil media as well, increasing support to its generality, as would be expected from mechanisms guided by thermodynamic processes.

The mechanism of seedling biomass (M) production cannot be assumed to be independent of absorbed metal concentration ($C_p$) because of the metal's biochemical toxicity. Fig. 6 shows the variation of M with $C_p$ together with the curve-fitted (SIIGMAPLOT 8.0, [18]) plot (broken lines) of the power-law function given by eqn 6.

$$M = \alpha C_p^{\beta} \qquad (6)$$

Where for DARK, 9H and 24H the $\alpha$ and $\beta$ values are, {0.0459, -0.5355}, {0.1276, -0.5381} and {4.6725, -1.1999} with the corresponding $R^2$ values for the fits at 95% confidence level as 0.92, 0.99 and 0.97 respectively.

DARK and 9H both seem to have a near inverse square root dependence on concentration but with different amplitude factors. It appears that the physiology of biomass production with respect to plant Ni up-take could be the same in both, with the presence of light energy (photosynthesis) contributing further to biomass in 9H by the increase in the amplitude factor. For 24H, high DLQ contributes to a high amplitude factor but the mechanism of biomass development under metal stress seems to be much more sensitive to metal concentration ( $|\beta| = 1.2$) confirming the previously observed DLQ enhancement of metal stress at the whole plant level. Although the present study concerns agar gel media, if soluble metal is available in the soil for absorption by the root, the same physical mechanism underlying eqn. 6 would apply to *B. juncea* growing under the same photo period conditions. We infer that strategies to raise absorbed metal concentrations such as by using organic acids [14] would likely result in reduced plant biomass.

The consequences of the interdependence of biomass and plant metal concentration on the mass of Ni absorbed, $M_{Ni}$, is seen in fig. 7. Since mechanistically, the response of DARK and 9H relevant to metal absorption appear to be the same, they show a similar trend – inverse dependence of $M_{Ni}$ with M whereas 24H shows a small increase of $M_{Ni}$ with M. Using eqns 1 and 6,

$$M_{Ni} = (1/\alpha)^{1/\beta} \cdot M^{(1/\beta + 1)} \qquad (7)$$

Substituting the values of $\alpha$ and $\beta$ for the 3 cases,

$$M_{Ni,\ DARK} = 2.116\text{x}10^{-3} / M$$

$$M_{Ni,\ 9H} = 1.628 \text{ x } 10^{-2} / M$$

$$M_{Ni,\ 24H} = 3.613\ (M^{1/6}) \qquad (8)$$

The amplitude factors in eqn. 8 bear a direct correlation to the amount of photosynthesis taking place, with the lowest for DARK and the highest for 24H. At low biomasses (M) the mass of Ni absorbed will be the highest for the 9H and low for 24H but lowest of all for DARK even though it absorbs the highest concentrations of Ni, because of its very low amplitude factor. This is corroborated in fig. 7 by the curves calculated using eqn. 8 and the actual experimental masses. Low M implies high $C_p$ (eqn. 6) which in turn implies high $C_s$ or soil metal concentration (eqn. 5/ fig. 5). Thus for plants that follow a physiological metal stress response similar to 9H, $M_{Ni}$ can be maximized at high soil metal concentrations below the fatally toxic limit. $M_{Ni}$ increases very slowly with M in the 24H case (eqn. 8/ fig. 7) so that plants that like it, grow rapidly (by photosynthesis), will be able to increase $M_{Ni}$ at low absorbed metal concentrations (eqn. 6). This implies growth in media (soil) with low metal content (eqn.5). From eqn. 8 we see that the functional form of 9H's physiological response dominates over the others, therefore the possibility for maximizing $M_{Ni}$ would appear to be in plants that follow a similar physiological metal stress response as 9H *B.juncea*, i.e. moderate photosynthetic growth at moderate temperature and relative humidity. This last observation might throw some light on the apparent correspondence between low growth rate and high metal accumulation in many hyper-accumulators although detailed studies aimed specifically at such plants are required for definite answers. For *B. juncea* at least, a frequent choice in phytoremediation schemes [3,14], the strong trade-off between M and $C_p$ for solubulized toxic metals, may mean that only a certain zone of plant toxic metal concentration and plant biomass may allow the maximal phytoremediation efficiency with extremes of either factor causing a net decrement. It would be interesting to see what kind of trade-off exists in transgenic *B. juncea* [3] modified to enhance metal up-take.

Perhaps not so surprising is the similarity of eqn. 8 with allometric scaling equations of plant metabolism that relate various metabolic dependent observables in a power-law manner to the plant mass [21]. The ability to accumulate metals requires metabolic energy (active processes) and that should in some way obey the universal power –law manifestation. From eqn. 8 we see that the power-law exponent vis à vis the Ni absorption capability of the plant *B. juncea* depends on its metabolic state, in this case as dictated by the ability to photosynthesize (i.e. DLQ), so that it varies from -1 to 1/6.

While field conditions represent a far greater complexity than agar gel media, if the metal is available to the plant root for absorption in soluble form and if the plant is not stressed in other ways, the physical processes and associated principles underlying eqns. 5-8 would still apply and the same conclusions would hold. It is also instructive to note that in the field, variation in the DLQ is naturally brought about by the seasons and near 24h daylight is possible at extreme latitudes at certain times of the year.

## 4 Conclusions

The factorial growth experiment incorporating the concentration of Ni in the medium, $C_s$, and the diurnal light quanta, DLQ, as independently variable parameters, shows that rate of photosynthetic driven growth impacts the capacity of the *B. juncea* seedling to absorb Ni. Light enhancement of Ni toxicity is evidenced by increased stunting of root lengths and a steeper decrease of total biomass, M, at high $C_s$. The shoot mass index (SMI) was modeled to allow a non-destructive estimate and its variation with $C_s$ indicates possibly that the plant may try to increase sequestration volume in the shoot in response to high metal intrusion. The analysis of the relationship between the absorbed metal concentration $C_p$ and $C_s$ and between $C_p$ and M shows that $C_p$ is not independent of $C_s$ but rather that it varies logarithmically with it, and that M depends strongly on $C_p$ especially for fast photosynthetically driven growth of the plant. The mathematical functions relating $C_s$, $C_p$ and M seem to conform to the underlying thermodynamics of the system [19] and the recent discussions of the universal power-law scaling of plant metabolic properties with mass [21]. The sum total of these observations indicate that for the metal accumulator *B. juncea* with regard to its capacity to accumulate Ni, the overall metabolic nature of the plant is important – neither rapid biomass increase nor a high metal concentration capability favored the removal of high metal mass from the medium, but rather the plant with the moderate photosynthetically driven biomass growth and moderate metal concentrations demonstrated the ability to remove the maximum mass of metal from the medium in the same growth period.

**Acknowledgements**: We would like to thank Mr. Jonathan Braisted and the staff at the University of Texas Nuclear Engineering Teaching Lab for helping us with the neutron activation portion of this project. Funding for this work was made possible by the grants from the NSF, NSF C-RUI number 0330815, and from the CONACYT, project number 12445.

## Figure Captions

**Fig. 1**: The temporal variation of the root lengths of *B. juncea* seedlings, grown in 1/2XMS agar media treated with 0 and 50 μM nickel acetate tetrahydrate solution and subjected to the variable diurnal light quanta of 0 (DARK), $4.32 \times 10^6$ (9H) and $1.20 \times 10^7$ (24H), μmol of photons.$m^{-2}.d^{-1}$. Error bars on the data points have been omitted for the sake of visual clarity. Lines through the points are eye-guides only.

**Fig.2**: The same as in fig. 1 except that the Ni(II) treatments are at the concentrations of 100, 150 and 200 μM.

**Fig. 3**: The variation of the total dry mass of the seedling at the 3 different light quanta treatments (DARK, 9H and 24H) and with Ni concentrations of 0, 50, 100, 150 and 200 μM in the media. The lines through the points are eye-guides only.

**Fig. 4**: The same as in fig. 3 except that the y-axis gives the shoot mass index (eqn. 4).

**Fig. 5**: The variation of the concentration of Ni absorbed by the plant, $C_p$, as a function of the concentration of Ni in the medium, $C_s$, at the three different diurnal light quanta. The broken lines are the curve-fitted calculated values (eqn. 5). The open squares are the data of Salt et al [12] and the broken line passing through them, is the curve-fit to eqn.5.

**Fig. 6**: The variation of seedling biomass, M, with Cp at the three diurnal light quanta. The broken lines are the curve-fitted calculated values (eqn. 6)

**Fig. 7**: The variation of mass of Ni absorbed by the *B. juncea* seedling, $M_{Ni}$, with M at the three different diurnal light quanta. The broken lines are the calculated values (eqn. 8).

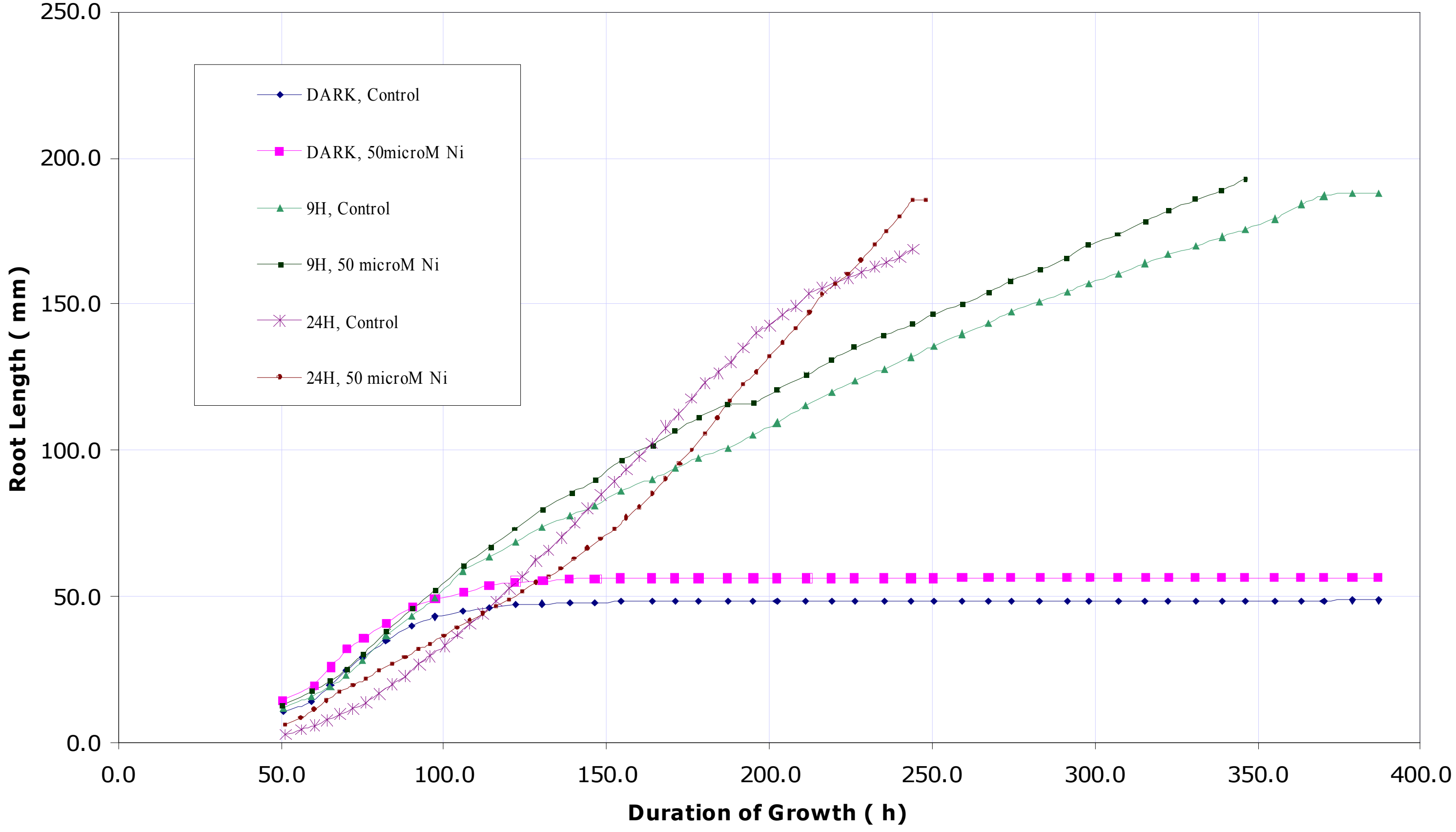

DARK, Control
DARK, 50microM Ni
9H, Control
9H, 50 microM Ni
24H, Control
24H, 50 microM Ni
Root Length ( mm)
Duration of Growth ( h)
0.0
50.0
100.0
150.0
200.0
250.0
300.0
350.0
400.0

Fig. 1

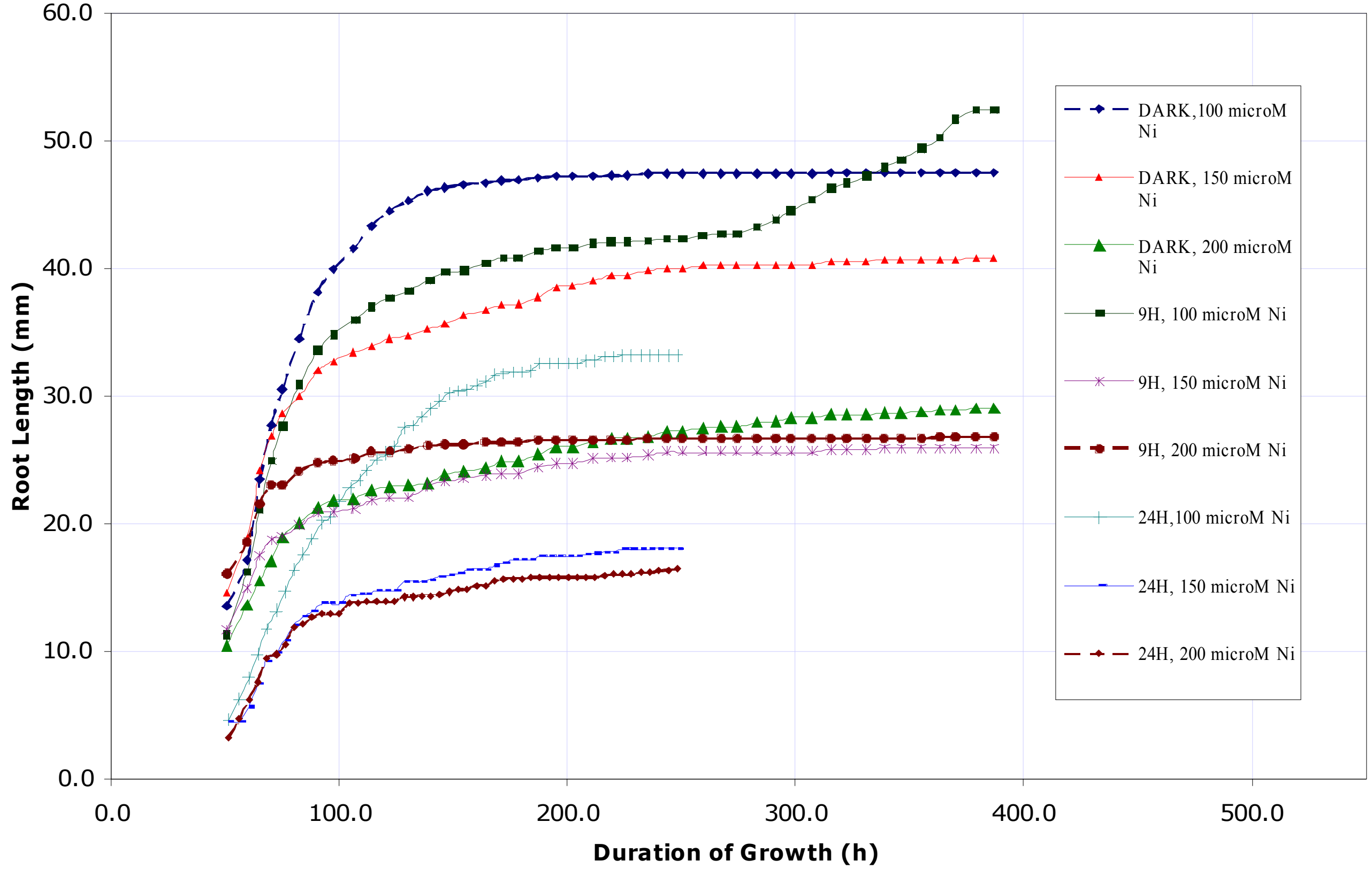

60.0
50.0
40.0
30.0
20.0
10.0
0.0
Root Length (mm)
0.0
100.0
200.0
300.0
400.0
500.0
Duration of Growth (h)
DARK,100 microM Ni
DARK, 150 microM Ni
DARK, 200 microM Ni
9H, 100 microM Ni
9H, 150 microM Ni
9H, 200 microM Ni
24H,100 microM Ni
24H, 150 microM Ni
24H, 200 microM Ni

Fig. 2

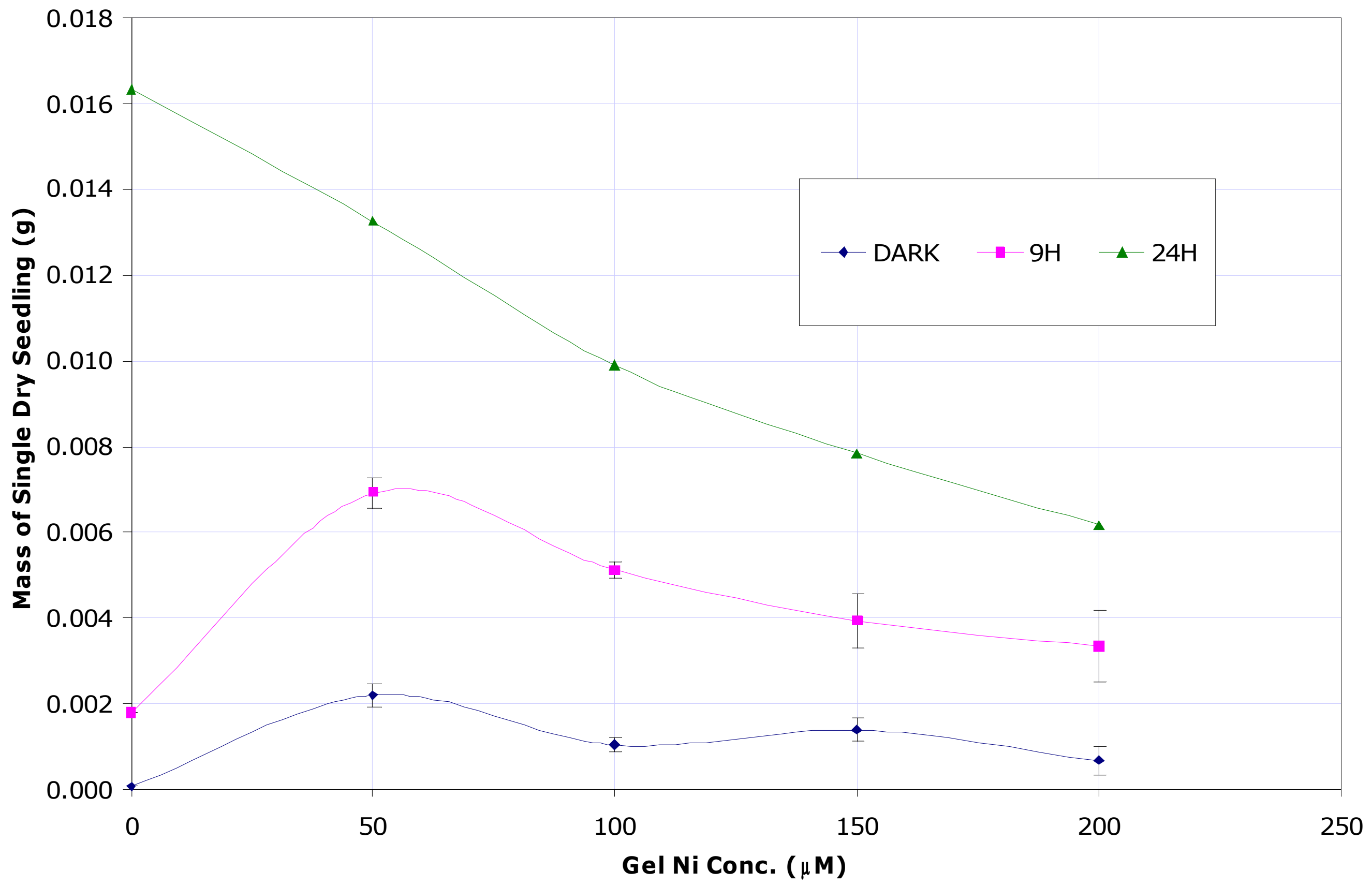

Fig. 3

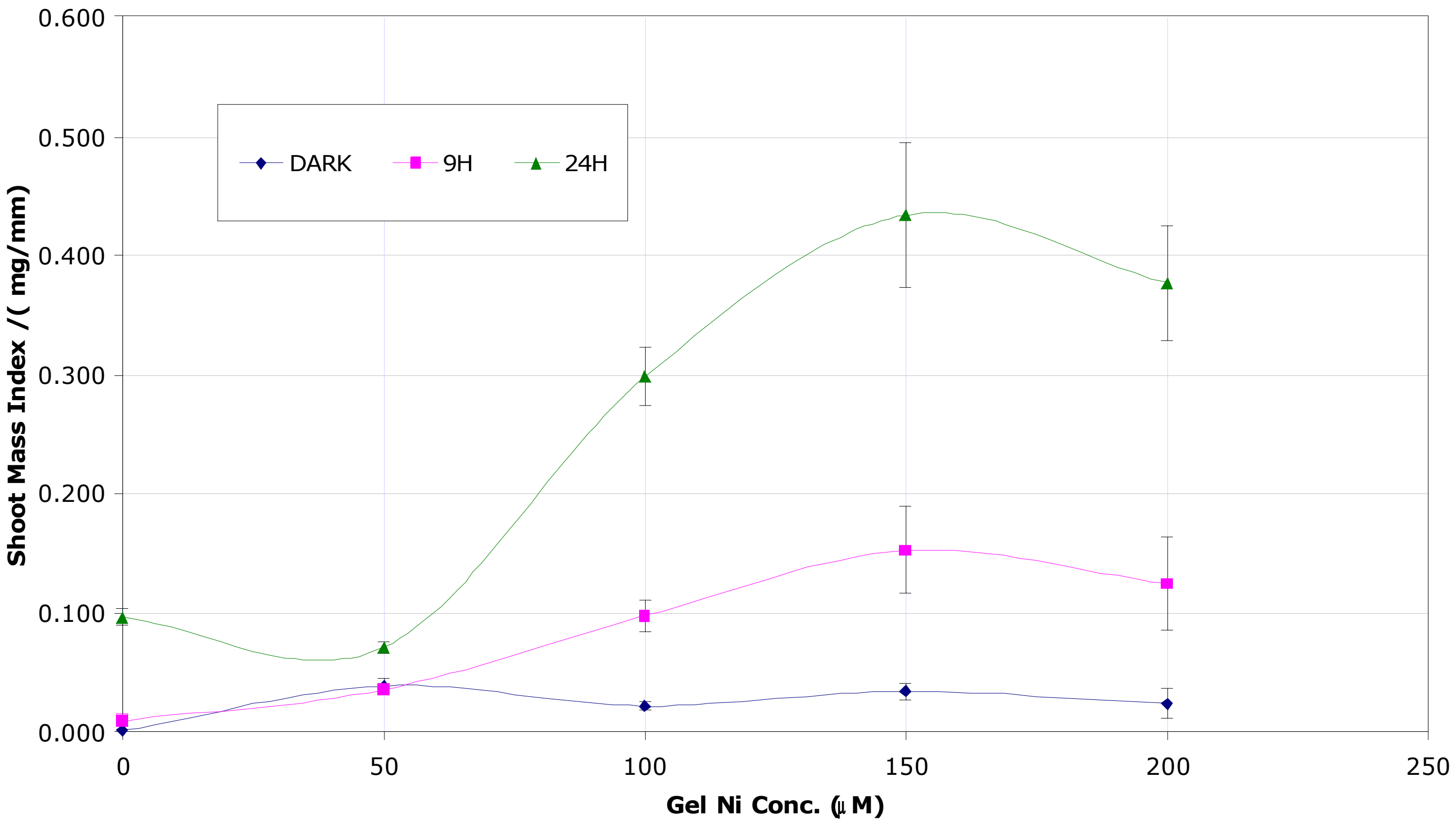

Fig. 4

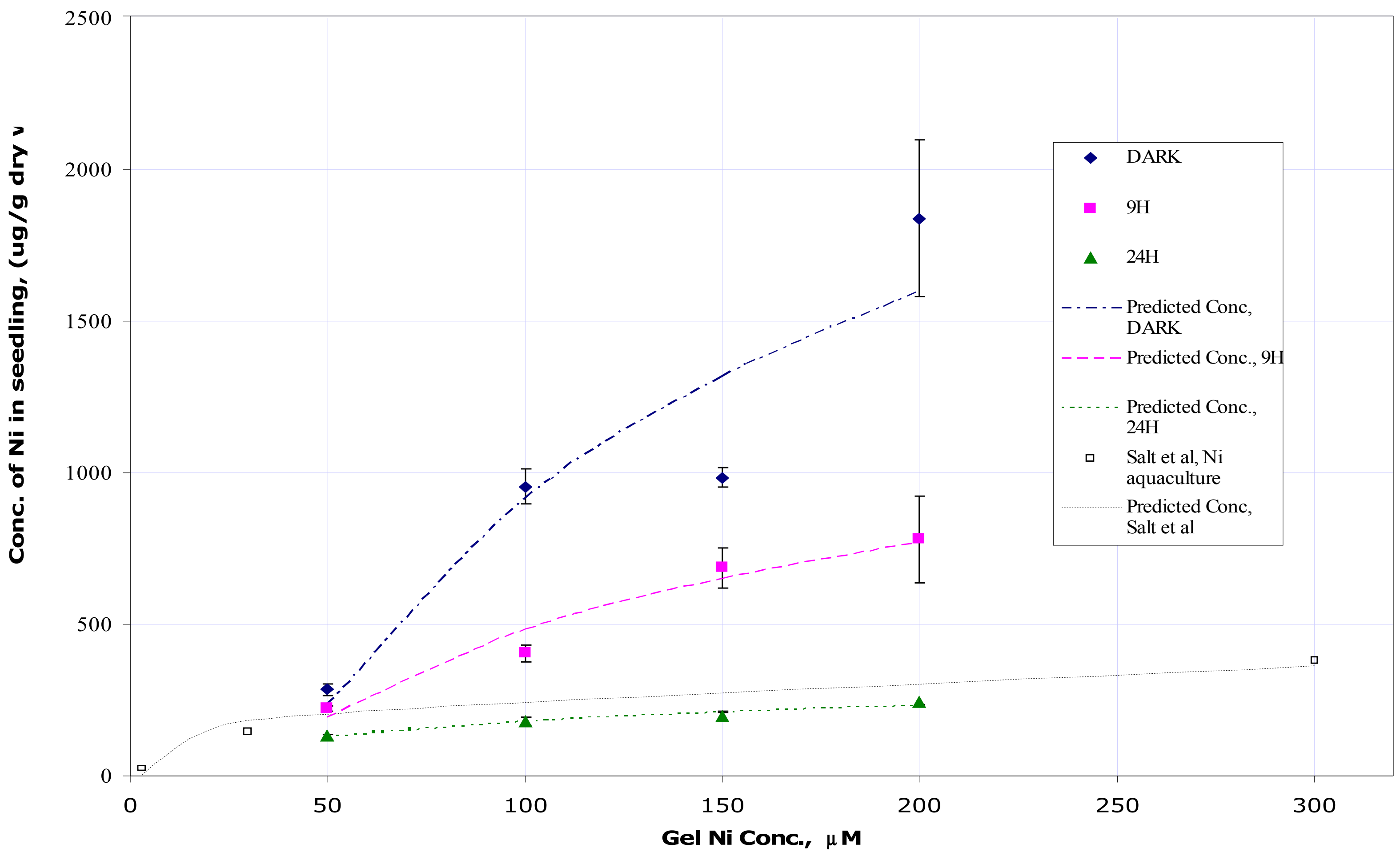

2500
2000
1500
1000
500
0
0
50
100
150
200
250
300
Conc. of Ni in seedling, (ug/g dry v
Gel Ni Conc., µM
DARK
9H
24H
Predicted Conc, DARK
Predicted Conc., 9H
Predicted Conc., 24H
Salt et al, Ni aquaculture
Predicted Conc, Salt et al

Fig. 5

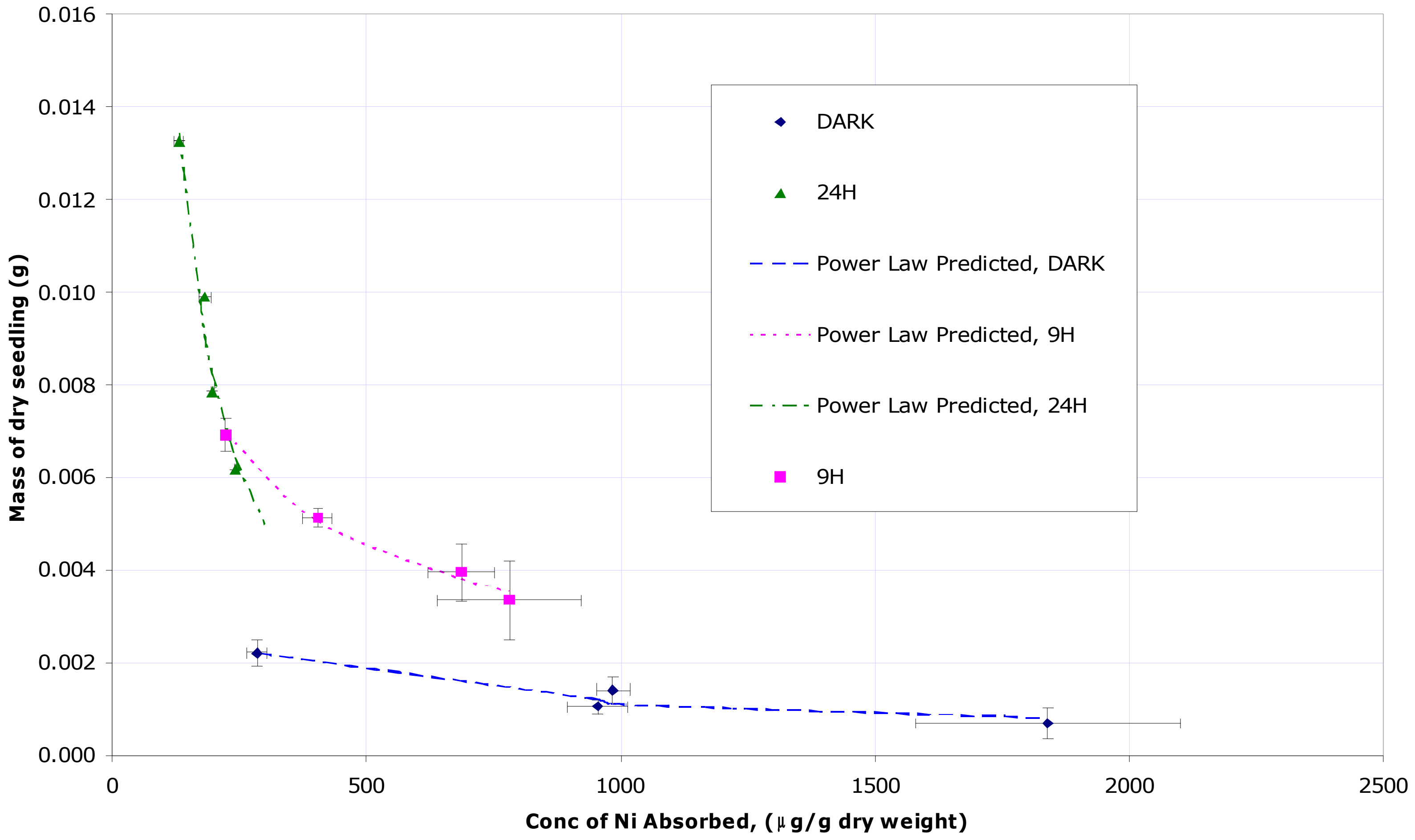

0.016
0.014
0.012
0.010
0.008
0.006
0.004
0.002
0.000
Mass of dry seedling (g)
0
500
1000
1500
2000
2500
Conc of Ni Absorbed, (µg/g dry weight)
DARK
24H
Power Law Predicted, DARK
Power Law Predicted, 9H
Power Law Predicted, 24H
9H

Fig. 6

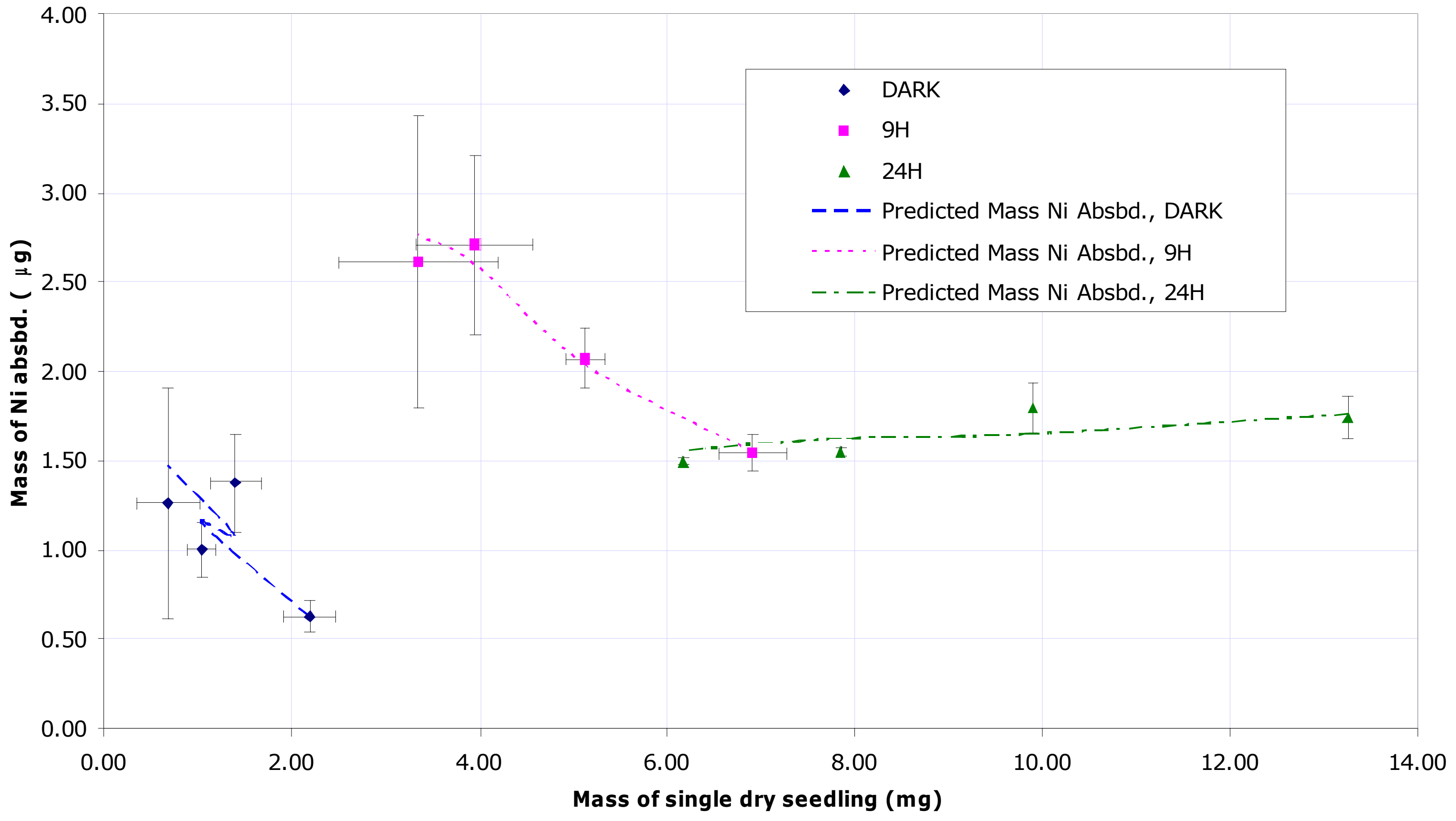

Fig. 7

Short introductory text for the "Table of Contents" of the ISEB Leipzig 2006 conference proceedings of the paper, 'The light quanta modulated physiological response of Brassica juncea seedlings subjected to Ni(II) stress' by N. Dasgupta-Schubert et al.

Interesting insights are presented into new possibilities of phytoremediation. The whole plant physiological response of *B. juncea* subjected to Ni stress, is investigated by a factorial growth experiment where the gel Ni concentration and plant biomass are independently varied, the latter by the photosynthetic light quantum.